\documentclass[12pt]{article}
\usepackage{epsfig}

\begin{document}

\title{Structure functions generated by zero sound excitations}
\author{V. A. Sadovnikova (St. Petersburg, INP)
}
\date{}
\maketitle

\begin{abstract}
In isospin symmetric and asymmetric nuclear matter we study the structure functions  connected with  the zero sound excitations. In the simple model with Landau-Migdal quasiparticle interactions the density response $\Pi^R(\omega,k)$  of the nuclear matter to the small isovector external  field $V_0(\omega,k)$ is considered.
In the previous papers the complex branches of solutions $\omega_{si}(k)$, $i=n,p,np$ of the dispersion equation corresponding to zero sound excitations in the symmetric and asymmetric nuclear matter are obtained. Now we investigate the structure and response functions of nuclear matter based on these solutions.
\end{abstract}

\section{Introduction}
In this work, the linear response and the corresponding structure functions of the nuclear matter to isovector external field \cite{AGD,Lip} are  studied.
The main results  concern  the different branches of zero-sound excitations in symmetric (SNM) and asymmetric (ANM) nuclear matter and response of matter to the external field exciting  these collective states.

There are a lot of publications describing  the different types of the excited collective states. In paper \cite{LipPe} the three types  of excitations are presented in the framework of local isospin density approximation, their contributions to the energy-weighted sum rules  in nuclei are evaluated. The response functions obtained in the article are used to compute  the frequency  and strength of the collective modes for different asymmetry parameter and density of matter.

In the papers  \cite{Moraw1,Moraw2} the energies and widths of the giant excitations and the corresponded strength functions are considered in the asymmetric nuclear matter and in nuclei.  The zero-sound dispersion equation was constructed on the basis of the non-Markovian kinetic equation. Two (isoscalar and isovector) complex modes are obtained and a new mode predicted which exists in ANM only.  The method of \cite{Moraw1,Moraw2} permits the transparent  addition of physical processes that describe  the resonance damping.

In \cite{3-co} propagation of the sound modes is considered  on the basis of the kinetic theory including collisions, temperature and memory effects. In \cite{4-co}   results for zero sound in nuclear matter considered in the framework of Landau-Migdal theory, are applied to the giant resonances in nuclei. To obtain the width of resonances and its temperature dependence the developed   kinetic theory was used.

A convenient model was suggested in  \cite{BV} to investigate the role of the effective nucleon-nucleon interaction in description of the giant resonances in the hot nuclei and  dependence of their energies and widths on the temperature.

In \cite{6-co}  a linear response theory starting from a relativistic kinetic equations is developed within a Quantum-Hadro-Dynamics effective field picture of the hadronic phase of nuclear matter. The dispersion relations are derived, they give the sound phase velocity and the internal structure of the normal collective modes, stable and unstable.

The detailed review of the linear response theories was presented in \cite{PDN} with the following study of the response to isospin flip probe \cite{1905}. In our paper we follow the main directions of \cite{PDN}.
We consider the density response function $\Pi^R(\omega,k)$ of the  medium (nuclear matter) to the weak external  field $V_0(t,r) = \lambda \tau_{3} e^{i\vec q\vec r-i(\omega+i\eta)t}$ \cite{AGD,Lip}.  The form of the response function we consider as determined by the zero sounds excitations of the  nuclear matter.
Structure function is defined as \cite{AGD,Lip}
\begin{equation}\label{1}
S(\omega,k) = -\frac1\pi{\rm Im} \Pi^R(\omega,k).
\end{equation}

In Sect.~2 we present the dispersion equation for zero sound excitations and recall its connection with response functions. In Sect.~3 the branches of zero sound frequencies in SNM and ANM are presented and in Sect.~4 the response functions are expressed as a sum over the different channels of collective state damping (connected  with the different branches). Results are discussed in Sect.5.

\section{Response functions}
We use the effective Landau-Migdal interaction between the quasiparticles \cite{MZL}
\begin{equation}\label{19}
{\cal F}(\vec\sigma_1,\vec \tau_1;\vec\sigma_2,\vec \tau_2) = C_0\left(F + F'(\vec\tau_1\vec\tau_2) + G(\vec\sigma_1\vec\sigma_2)
 + G' (\vec\tau_1\vec\tau_2)\, (\vec\sigma_1\vec\sigma_2)\right),
\end{equation}
where $\vec\sigma$, $\vec \tau$ are the Pauli matrices in the spin and isospin spaces. $C_0= N^{-1} =\frac{\pi^2}{p_0\,m_0}$ where $N$ is the density of states of one sort of particles, $m_0=0.94$GeV.

We consider  the response of ANM to the isovector external field $V_0(\omega,k)$. Below we follow the paper \cite{1402}. In this case the  response function $\Pi(\omega,k)$ (retarded polarization operator \cite{Lip}) can be presented as \cite{Lip,HNP1}
\begin{equation}\label{6r}
\Pi(\omega,k)= \Pi^{pp}(\omega,k) + \Pi^{nn}(\omega,k) - \Pi^{pn}(\omega,k) - \Pi^{np}(\omega,k).
\end{equation}
Here $\Pi^{\tau\tau'}(\omega,k)$ are the RPA $ph$ propagators averaging over momentum to get the response functions, interconnected by the Bethe-Salpeter equations. These equations for $\Pi^{pp}(\omega,k)$ and $\Pi^{np}(\omega,k)$ are
\begin{eqnarray}
&& \Pi^{pp} = A^p + A^p\,F^{pp}\,\Pi^{pp} + A^p\,F^{pn}\,\Pi^{np},
\nonumber\\
&& \Pi^{np} = \quad \quad    A^n\,F^{np}\,\Pi^{pp} + A^n\,F^{nn}\,  \Pi^{np} .
\label{7}
\end{eqnarray}
Here the following designations are used
\begin{equation}\label{3r}
F^{pp}=  C_{0p}\,(F + F'), \quad F^{nn}=  C_{0n}\,(F + F'), \quad  F^{pn} = F^{np} = (C_{0p}\,C_{0n})^{1/2}(F - F').
\end{equation}
Here $C_{0p}= N_p^{-1}=\frac{\pi^2}{m_0\,p_{Fp}}$, $C_{0n}= N_n^{-1} = \frac{\pi^2}{m_0\,p_{Fn}}$.
Functions $A^\tau = A^\tau(\omega,k)$ are the integrals over the particle-hole loops with  isospin  $\tau$ \cite{MZL,cc} and, on the other hand, they are the response to the external field of the matter consisting of the non-integrated particles, $\Pi_0^{\tau\tau'}$. $A^\tau(\omega,k)$ have the form of the Lindhardt functions \cite{Lip,cc} (Appendix).

The system Eq.(\ref{7}) we rewrite using the matrix ${\cal M}$:
\begin{equation}\label{74}
{\cal M} =
\left(
\begin{array}{cc}
(1 - A^p\,F^{pp}) & - A^p\,F^{pn}\nonumber\\
- A^n\,F^{np} & (1 - A^n\,F^{nn})\nonumber\\
\end{array}
\right).
\end{equation}
So, for two pairs $\Pi^{pp}$, $\Pi^{np}$ and $\Pi^{nn}$, $\Pi^{pn}$ two systems can be written:
\begin{equation}\label{75}
{\cal M}
\left(
\begin{array}{c}
\Pi^{pp}\nonumber\\
\Pi^{pn}\nonumber\\
\end{array}
\right)
=
\left(
\begin{array}{c}
A^{p}_0\nonumber\\
0 \nonumber\\
\end{array}
\right),
\quad
{\cal M}
\left(
\begin{array}{c}
\Pi^{nn}\nonumber\\
\Pi^{np}\nonumber\\
\end{array}
\right)
=
\left(
\begin{array}{c}
A^n\nonumber\\
0 \nonumber\\
\end{array}
\right).
\end{equation}

Solving  the systems (\ref{75}) we obtain $\Pi^{\tau\tau'}$ as
\begin{equation}\label{81}
\Pi^{nn}= \frac{A^n(1-A^p\,F^{pp})}{\det{\cal M}} \equiv \frac{D^{nn}}{\det{\cal M}}, \quad \Pi^{pn} =  \frac{A^n\,A^p\,F^{pn}}{\det{\cal M}} \equiv \frac{D^{pn}}{\det{\cal M}}.
\end{equation}
 Changing $n \leftrightarrow p$ we obtain $\Pi^{pp}$ and $\Pi^{np}$. Denominator $\det{\cal M}$ is the same in all terms.
Below we calculate $\Pi^{\tau\tau'}$ as singular functions with poles in the zeros of the denominator.

The full isovector ($iv$) retarded polarization operator is the sum (Eq.(\ref{6r}))
\begin{equation}\label{6pr}
\Pi= \Pi^{pp} + \Pi^{nn} - \Pi^{pn} - \Pi^{np}= \frac{(D^{pp} + D^{nn} - D^{pn} - D^{np})}{\det{\cal M}}
\equiv \frac{D^{iv}(\omega,k)}{E(\omega,k)},
\end{equation}
here $\det{\cal M}(\omega,k)$ is denoted as $E(\omega,k)$ and $D^{iv}=(D^{pp} + D^{nn} - D^{pn} - D^{np})$.
Dispersion equation for the frequencies of zero sound excitations of systems is $E(\omega,k)= 0$. In detailed form it is
\begin{equation}\label{5c}
E(\omega,k) = 1 - C_{0p}(F+F')A^p(\omega,k) - C_{0n}(F+F')A^n(\omega,k) + 4F F' C_{0p}C_{0n} A^p(\omega,k)A^n(\omega,k)= 0.
\end{equation}
This dispersion equation has  some types of solutions, i.e., there are some frequencies $\omega_i$ at every $k$ when Eq.(\ref{5c}) is satisfied. Solutions are presented as the function of the frequency of excitations depending on the wave vector: the branches $\omega_{si}(k)$. At some conditions the solutions are complex with the different  physical sense of the imaginary parts. The  imaginary part of $\omega_{si}(k)$ corresponds to the definite channel of excitation damping.

\subsection{Response functions in symmetric nuclear matter (SNM)}
In SNM  the dispersion equation Eq.~(\ref{5c}) can be presented as a product
\begin{equation}\label{4a}
E(\omega,k) = (1-F^{nn}A^n(\omega,k))\,(1-F^{pp}A^p(\omega,k))-(A^p(\omega,k) F^{pn})\,(A^n(\omega,k) F^{np}) =
\end{equation}
$$
= (1 - C_0 F A(\omega,k))\,(1 - C_0 F' A(\omega,k)),
$$
here $A = A_n+A_p$, $C_{0n} = C_{0p} =C_{0}$. The factorization  means that  there are two sorts of solutions (isoscalar and isovector) to Eq.(\ref{4a}). The isoscalar collective mode appears due to isoscalar quasiparticles interaction $F$, Eq.(\ref{19}). At every $k$ it is solution  of dispersion equation $1 - C_0 F A(\omega,k)=0$,  the first factor in Eq.(\ref{4a}). The  isovector collective modes are the solutions  of the dispersion equation $1 - C_0 F' A(\omega,k)=0$.

Factorization permits to conclude that in SNM the isoscalar  and isovector collective excitations do not interact.
This factorization is broken at $\beta\neq 0$.  If the  different values of $F$ and $F'$ are used then the isovector and isoscalar branches differ significantly. And we can trace  and separate them when $\beta$ is changed.

The isoscalar ($is$) response function to the isoscalar  external field, instead of Eq.(\ref{6pr}), is \cite{HNP1}:
\begin{equation}\label{7pr}
\Pi= \Pi^{pp} + \Pi^{nn} + \Pi^{pn} + \Pi^{np}.
\end{equation}

For the response functions we have in SNM:
$$
\mbox{isovector:  } \Pi(\omega,k)= \frac{A}{(1 - C_0 F' A)},\quad
\mbox{isoscalar:  } \Pi(\omega,k)= \frac{A}{(1 - C_0 F A)}.
$$

The branches of solutions are very sensitive to the parameters of medium and quasiparticle interaction \cite{vs14}. Results of this paper are obtained and presented in isovector modes with $F'=1.0, F=0$, Eq.(\ref{19}). Then in SNM we have two branches of solutions $\omega_{s}(k)$ and $\omega_{s1}(k)$ \cite{cc} which are used below.

\section{Solutions of the dispersion equation}
We present the results for isovector structure functions in ANM with the asymmetry parameter $\beta=(\rho_n-\rho_p)/(\rho_n+\rho_p)=0.2$.
In \cite{cc} the complex branches of zero sound excitations $\omega_{si}(k)$, $i=n,p,np$ were obtained and demonstrated for different $\beta$.
Dispersion equation with  isovector  $F'$ Landau-Migdal quasiparticle interaction (\ref{19}) is presented in Eq.(\ref{5c}).  Now the contributions of these solutions into the response and strength function are studied. In nuclei, the imaginary parts of these solutions correspond to the widths of resonance decays. When $i=p$ we have the branch $\omega_{sp}(k)$, in nuclei the width appears  due to damping of excitations by emission of the proton; $i=n$, $\omega_{sn}(k)$ is damping by emission of neutron. When  $i=np$, the imaginary part of  $\omega_{snp}(k)$ is due to emission of the nucleon without the fixed isospin. Every of these processes correspond to the separate channel of decay. In nuclear matters, the width can be explained as the departure of the part of $ph$ pairs from the process formation of the collective mode and transition to the state of non-interacting pairs.

We consider the dispersion equation Eq.(\ref{5c}) on the complex $\omega$-plane \cite{cc}. Functions $A^\tau(\omega,k)$ are the Lindhardt functions which contain the logarithms (Appendix). These logarithmic functions have the cuts and it is possible to transit to unphysical sheets through these cuts. The points on the  cuts correspond to the energies of the non-interacting $ph$ pairs.

At small $k$ solutions of Eq.(\ref{5c}) are real and are placed on the real axis out of cuts. When $k$ increase the overlapping of real solutions and the cuts  appeared. We look for and find the complex solutions on the nearest unphysical sheet of the logarithmic Riemann surface. The imaginary part of solution is determined  by the nature of cut under which we look for this solution of Eq.(\ref{5c}). When the continuation of the branch of solutions is placed on the unphysical sheet under the cut, we say that cut is open.  Suppose, this is the neutron cut corresponding to the energies of the free neutron $ph$ pairs.  In nuclei the imaginary part of solution placed under this neutron cut corresponds to damping of excitations due to  the emission of the neutron. Technically, $A^n(\omega,k)$, Eq.(\ref{a1}), is taken on the unphysical sheet of logarithm, but $A^n(-\omega,k)$, $A^p(\omega,k)$, $A^p(-\omega,k)$ are on the physical sheet of $\omega$-plane. Their cuts are considered as closed.

Eq.(\ref{5c}) has the cuts of two types \cite{cc}: corresponding to logarithmic cuts of $A^p$ and of $A^n$.
 The cuts may be closed or open. We obtain some version of solutions.
1) There is a stable real solution (no channel is open). This solution belongs to $\omega_s(k)$ when $Z=N$, to $\omega_{sn}(k)$ when $N>Z$ and to $\omega_{sp}(k)$ when $Z>N$.
2) The proton channel is  open. Neutron channel is closed. Solutions are complex. They correspond to the excitations damping due to emission of proton. Solutions belong to $\omega_{sp}(k)$.
3) The neutron channel is open. The proton channel is closed. Solutions belong to $\omega_{sn}(k)$.
4) The both  channels are open. In this case one nucleon is emitted but this approach does not fix its isospin. Solutions belong to $\omega_{snp}(k)$. This complex branch presents the continuation of $\omega_{s}(k)$ (obtained in SNM) into ANM.

We do not have the complex physical solutions (only the real ones) on the physical sheet of the complex $\omega$-plane. We consider the solutions as a physical ones if there is an analytical continuation to them over the some parameter from the real physical solutions.

\section{Expression for the structure functions $S(\omega,k)$}
When we consider scattering of the photons on the nuclei, the different channels ($l$) are separated:  $(\gamma,p)$, $(\gamma,n)$.
In our simple model there are two mechanisms which give the contribution in $S(\omega,k)$ and, hence, in the cross section of reactions. Let consider $(\gamma,n)$ reaction: first mechanism is the excitation of the collective state $\omega_{sn}(k)$  and its further damping due to emission of neutron (if ${\rm Im}\omega_{sn}(k)\neq 0)$ and second mechanism is the direct knock out of neutron, described by ${\rm Im}A^n$. The same with reaction $(\gamma,p)$.  But the contribution of $\omega_{snp}(k)$ is impossible to  connect with a definite reaction, so we attribute it to a separate channel (and a separate peak in figures).

The cross sections in these channels are proportional to the strength functions $S_l(\omega,k)$. The total strength function  $S(\omega,k)$ is a sum over the channels
\begin{equation} \label{7r}
 S(\omega,k) = \sum_{l} S_l(\omega,k),
 \end{equation}
 where $l=n,p,np$. We can rewrite  $\Pi(\omega,k)$ Eq.~(\ref{6pr}) as a sum over the poles.  At the beginning we present the inverse determinant as a sum over the poles
 \begin{equation} \label{8dec}
\frac{1}{E(\omega,k)} = \sum_l\left(\frac{R_l(\omega_{sl},k)}{\omega-\omega_{sl}(k)} + Reg_{l}(\omega,k) \right).
\end{equation}
 Here $Reg_l(\omega,k)$ is a smooth function near the pole. Residues $R_l(\omega_{sl},k)$ in the poles are calculated on the same unphysical sheets where the poles are placed
 \begin{equation} \label{9S}
 R_l(\omega_{sl},k)=\frac{1}{E'(\omega_{sl}(k))}= \frac{{\rm Re}\,(E') - I\,{\rm Im}(E')}{|E'|^2}, \mbox{ where }
E'(\omega_{sl}(k))= \frac{dE(\omega,k)}{d\omega}|_{\omega\to \omega_{sl}(k)}.
\end{equation}
Then polarization operator has the form
 \begin{equation} \label{10S}
\Pi(\omega,k)= \sum_l D^{iv}(\omega,k)\left(\frac{R_l(\omega_{sl},k)}{\omega-\omega_{sl}(k)} + Reg_{l}(\omega,k) \right).
\end{equation}
The structure function $S(\omega,k)$ is equal
\begin{equation} \label{11S}
 S(\omega,k) = \sum_{l} S_l(\omega,k) = -\frac{1}{\pi}{\rm Im}\sum_l D^{iv}(\omega,k)\left(\frac{R_l(\omega_{sl},k)}{\omega-\omega_{sl}(k)} + Reg_{l}(\omega,k) \right).
 \end{equation}

Now let define the envelope curve of the pole terms
\begin{equation} \label{14S}
S^e(\omega,k)= -\frac1\pi \Sigma_{\tau,\tau'} {\rm Im} \left[D^{\tau,\tau'}(\omega,k) \Sigma_i\,R^i(\omega_{si},k)/ (\omega-\omega_{si})\right].
\end{equation}

\subsection{Contribution of the free $ph$ pairs into $S(\omega,k)$}
 The second contribution in  $S(\omega,k)$ is related to the interaction of the external field with the non-interacting proton and neutron $ph$ pairs and it is included in $Reg_{l}(\omega,k)$ in Eq.(\ref{11S}). Response function ($S_{fr}(\omega,k)$) may be obtained using Eq.(\ref{6pr}) with $F'=0$. Then
\begin{equation} \label{15S}
S_{fr}(\omega,k)= -\frac1\pi {\rm Im} (A^p(\omega,k) + A^n(\omega,k)).
\end{equation}

\section{Results}
In this section we demonstrate  the branches of solutions $\omega_{si}(k)$ and the structure functions (\ref{14S}).  generated by these solutions.  Calculations are made in SNM and in ANM with asymmetry parameter $\beta=0.2$, at the equilibrium density $\rho= \rho_0= 0.17\,fm^{-3}$, $p_0=0.268$GeV. Parameters of quasiparticle interaction (\ref{19}) are taken as $F'$=1.0, $F=0$.  In calculations the quasiparticle effective mass was  used: $m^*=0.8\,m_0$.

\subsection{SNM}
In Fig.1 we present results in SNM. The branches of zero sound solutions $\omega_{s}(k)$ and $\omega_{s1}(k)$ are shown in Fig.1($left$). The structure functions corresponding to these branches are presented  in Fig.1($right$). In the left figure  $\omega_{s}(k)$ (solid curve) is shown, it is real at $k<k_t$ and the imaginary part of branch starts at $k=k_t$. Here $k_t=k_t(\beta=0)=0.34p_0$. ${\rm Im}(\omega_{s}(k))$ describe the damping of excitations due to emission of one nucleon and the isospin of this nucleon is not fixed, it may be proton or neutron. At $k>k_t$, $\omega_{s}(k)$ is placed on the unphysical  logarithmic sheets  associated with  the logarithm functions in the both $A^n(\omega,k)$ and $A^p(\omega,k)$ ($A^p=A^n$  in SNM).

The second branch $\omega_{s1}(k)$ is placed completely on the unphysical sheet of $A^p(\omega,k)$ or $A^n(\omega,k)$ but not the both \cite{cc}.
It starts at $k_c=0.52p_0$ and has an imaginary part considerably larger (by absolute value) than that of  $\omega_{s}(k)$. ${\rm Im}(\omega_{s1}(k))$ describes the damping of excitations due to emission of nucleons with one definite isospin (only protons or only neutrons). In ANM this branch is splitting into two branches $\omega_{sp}(k)$ and $\omega_{sn}(k)$ (Fig.2 ($left$)).
They are placed on the different unphysical sheets and have $\omega_{s1}(k)$ as a limit at $\beta\to 0$. This is true provided that $\omega_{s1}(k)$ exists, i.e., $k>k_c$ \cite{cc}.

In Fig.1($right$) the structure functions for $k/p_0$=0.2, 0.6 are shown. In Fig.1($left$) we see that only the real solution  at $k/p_0$=0.2 is found. So,  $S^e(\omega,k)$ has an infinite peak (solid curve with stars) in  Fig.1($right$). At $k/p_0$=0.6  there are two complex solutions in Fig.1($left$). So we see two wide maxima: the dotted one is generated by  $\omega_{s1}(k=0.6p_0)$, the solid one - by  $\omega_{s}(k=0.6p_0)$ (see Table).
Their widths are  determined by ${\rm Im}\,\omega_{s1}(k)$ and ${\rm Im}\,\omega_{s}(k)$.

\subsection{ANM}
In ANM  the branches, Fig.2($left$), and the corresponding structure functions, Fig.2($right$)  are presented. In the Fig.2($left$) we see $\omega_{sn}(k)$ (dotted curve), this branch starts at $k=0$ and it is real up to $k\leq k_t(\beta)$. For $0\leq k\leq k_t(\beta)$ the branch $\omega_{sn}(k)$ is continuation of  $\omega_{s}(k)$ (obtained in SNM) into ANM. $k_t/p_0(\beta=0.2)=0.19$ \cite{cc}. In Table the frequencies of the branches at $k/p_0=0.2, 0.6$ are presented in correspondence with Fig.2($left$).
At larger $k>k_t$, $\omega_{sn}(k)$ goes to unphysical sheet connected with $A^n(\omega,k)$ and starts to damp by emitting the neutrons.

The branch $\omega_{sp}(k)$ (dashed curve, in Fig.2($left$)) is placed on the  unphysical  sheet associated with $A^p(\omega,k)$ and is  damping by emission of protons. It appears at $\beta\neq0$ at the definite $k_p(\beta)$, see Eq.(\ref{a3}) in Appendix. The branch $\omega_{snp}(k)$ (dot-dashed curve) is placed on the  unphysical  sheet connected with both $A^p(\omega,k)$ and $A^n(\omega,k)$ and is  damping by emission of one nucleon.

In Fig.2($right$) we present structure functions for $k/p_0=0.2, 0.6$.
At $k/p_0=0.2$ there are a high narrow peak  which belongs to $\omega_{sn}(k)$ and a low wide peak relating to $\omega_{sp}(k)$ (solid with stars curves). Instead of one real solution  of Eq.(\ref{5c}) in SNM we obtain two complex solutions in ANM, so in the structure function instead of an infinite peak we obtain two maxima with a finite width.

At $k/p_0=0.6$ in Fig.2($right$)) there is a more rich form of the structure function with three maxima. There is  a dotted peak corresponding to $\omega_{sn}(k=0.6p_0)$.
 As well we see the dashed maximum generated by solution $\omega_{sp}(k=0.6p_0)$.   The maxima  of the dot-dashed curve corresponds to $\omega_{snp}(k=0.6p_0)$.   The branch $\omega_{snp}(k)$ is the continuation at $k>k_t(\beta=0)$  of $\omega_{s}(k)$  into ANM.

There are rather complicated associations between the branches of solutions in SNM and ANM \cite{cc}. This depends, first of all,  on value of $k$ for the fixed $\beta$. When $k<k_t$ and $\omega_{s}(k)$ is real it turn into $\omega_{sn}(k)$ at $\beta\neq0$ with $k$ increasing. When $k>k_t$ $\omega_{s}(k)$ is complex in SNM, it turn into the branch $\omega_{snp}(k)$ in ANM.

At $k/p_0=0.6$. As it was said at $k>k_c$ there is a splitting of $\omega_{s1}(k)$ into $\omega_{sn}(k)$ and $\omega_{sp}(k)$ in ANM.  We see that the maximum in Fig.1($right$) marked $'S1'$ is splitting  on the two maxima $1$ and $2$ in Fig.2($right$). The peak $'S'$ turn into maximum '3' in Fig.2($right$).

\subsection{Contributions of the free $ph$ pairs, Fig.~3}
In Fig.~2 the contributions of the pole terms in $S^e$ (\ref{14S}) was demonstrated and $S_{fr}(\omega,k)$ (\ref{15S}) was placed in term $Reg_l$. Now we show the both contributions in one figure and compare.
The envelope curve $S^e(Fig.3)$ is the sum of  $S^e$, (Fig.2($right$)), and $S_{fr}$.
\begin{equation} \label{16S}
S^e(Fig.3(left))= -\frac1\pi  {\rm Im} \left[D^{iv}(\omega,k) \Sigma_i\,R^i(\omega_{si},k)/ (\omega-\omega_{si})\right] + S_{fr}(\omega,k).
\end{equation}
In Fig.3($left$) the fat solid curve stands for the $S_{fr}$, Eq.(\ref{15S}). We see that  its value is comparable to contributions of poles. The form of the curve  results from the fact  that it is the sum of the proton and neutron ${\rm Im}A^\tau(\omega,k)$, Eq.(\ref{15S}), which have the maxima at the different energies due to different Fermi momenta $p_{Fp}$ and $p_{Fn}$.
  Fig.3 is made for $k/p_0=0.6$. The form of ${\rm Im}A(\omega,k)$ is presented in a lot of places, for example, in Figure\,3 \cite{PDN}.

We see that after adding of the free $ph$ terms, the two maxima formed  $S^e$ in Fig.2($left$) are conserved but the value of maxima is changed.

In Fig.3($right$) the contributions of $S_{fr}$ are regrouped in a such way that  the maximum associated with emission of protons  (pole and free $ph$ terms) was separated from that one connected with emission of neutrons. The proton and neutron maxima in structure function are increased and broadened after adding of the contribution of free $ph$ pairs.

\section{Summary}
The picture of the process described in this work is the following. The external field $V_0(\omega,k)$ excites the zero  sound collective states in nuclear matter. Frequencies of these states are obtained as solutions of the dispersion equation for zero sound, Eq.(\ref{5c}) \cite{cc}. The imaginary parts of solutions describe the damping of collective excitations due to emission of the nucleons (in the case of nuclei).

These branches $\omega_{si}(k)$ are the poles of response functions $\Pi^R(\omega,k)$.  We present the response of nuclear matter to external field through the sum over the poles. The every pole corresponds to the physical process: excitation of the stable state or excitation  of the collective state with the following decay   due to emission of definite particle. Then we present the  response function as a sum  of response functions for the separate processes.  In Fig.2($right$) the maximum of the  dotted curve correspond to the zero sound collective state excited by $V_0$ with the quantum numbers of this external field and damping due to emission of neutron. The maximum on the dashed curve correspond to the  damping due to emission of proton and the dot-dashed curve demonstrates the decay channel of an excited state according to the branch $\omega_{snp}$.

The external field $V_0$ excites the isoscalar-scalar  states in nuclear matter. It would be interesting to compare our results with the structure functions in nuclei. In \cite{vs14} there was a rather successful attempt to describe the frequencies of the  giant dipole resonances (GDR) using our  simple approach. The successive results for GDR were obtained in \cite{BV}.  But the  problem of our model at the transition to nuclei
 is the opposite order of proton and neutron maxima in $S(\omega.k)$. In Figs.~2 and~3 the maxima connected with the emission of protons are placed on the left  from neutron maxima. The reason is in the smaller $p_{Fp}$  in respect to $p_{Fn}$. But in the experiments the order is back, this can be explained by the Coulomb barrier for the emitted protons.

The simple model suggested in this paper permits to analyze  the structure functions separating the different processes which form $S(\omega,k)$.

\section{Appendix}
Functions $A^\tau = A^\tau(\omega,k)$ are the integrals over the particle-hole loops with  isospin  $\tau$ \cite{MZL,cc} and, on the other hand, they are the response to the external field of the matter with the non-integrating $ph$ pairs. $A^\tau(\omega,k)$ have the form of the Lindhardt functions \cite{Lip,cc}.
\begin{equation} \label{a1}
A^p = A^p(\omega,k) + A^p(-\omega,k), \quad A^n = A^n(\omega,k) + A^n(-\omega,k).
\end{equation}
and
\begin{equation} \label{a2}
A^{\tau}(\omega,k) =\ -2\frac1{4\pi^2}\
\frac{m^3}{k^3} \left[\frac{a^2-b_\tau^2}2 \ln\left(\frac{a+b_\tau} {a-b_\tau}\right)-ab_\tau \right],
\end{equation}
where $a=\omega-(\frac{k^2}{2m})$, $b_\tau =\frac{kp_{F\tau}}{m}$.
Functions $A^\tau$ have the  logarithmic cuts.
 \begin{equation} \label{a3}
 (1,1'): -\frac{kp_F}{m} + \frac{k^2}{2m}\ \le\ \omega\ \le\ \frac{kp_F}{m}+\frac{k^2}{2m}\ ,
  \quad  (2,2'):-\frac{kp_F}{m} - \frac{k^2}{2m}\ \le\  \omega\ \le\ \frac{kp_F}{m} - \frac{k^2}{2m}\ .
\end{equation}

In the text above it was mentioned that $\omega_{sp}(k)$ appears at $k=k_p(\beta)$.
 Calculations demonstrate that it exists at $\beta \neq 0$ and starts at  $k_p$  such that $\omega = \frac{k_pp_F}{m} - \frac{k_p^2}{2m}$ is the solution of the dispersion equation, Eq.(\ref{5c}). It means that the right point of the cut '2' is the solution and the branch $\omega_{sp}(k)$ appears in the point $2'$. We obtained $\omega_{sp}(k)$ at $k>k_p$ but at $k<k_p$ it is absent. Probably, the branch $\omega_{sp}(k)$ has the admixture of $2p-2h$ states at these $k$ and it is necessary other method of its description.

{}

\newpage

\begin{center}
{\bf Table.} Symmetric ($\beta=0$) and asymmetric ($\beta=0.2$) nuclear matter. The solutions $\omega_{si}(k)$ are presented in the form of $({\rm Re}\, \omega_{si}/p_0,\,{\rm Im}\,\omega_{si}/p_0)$ at $k/p_0$=0.2, 0.6.\\

\begin{tabular}{|c|c|c|}
\hline
  SNM & $k/p_0$=0.2          & $k/p_0$=0.6 \\
\hline
$\omega_{s}$ & (0.807$\cdot10^{-1}$,\,0.0) & (0.273,\,-0.790$\cdot10^{-2}$) \\
\hline
$\omega_{s1}$ &                      &(0.255, -0.226$\cdot10^{-1}$) \\
\hline
  ANM &&\\
\hline
$\omega_{sn}$ & (0.827$\cdot10^{-1}$,\,-0.700$\cdot10^{-4}$) & (0.279$\cdot10^{-1}$,\,-0.276$\cdot10^{-1}$) \\
\hline
$\omega_{sp}$ & (0.163$\cdot10^{-1}$,\,-0.720$\cdot10^{-2}$) & (0.229,\,-0.360$\cdot10^{-1}$) \\
\hline
$\omega_{snp}$ &                                           & (0.270,\,-0.851$\cdot10^{-2}$) \\
\hline
\end{tabular}
\end{center}

\newpage

\begin{figure}
\centering{\epsfig{figure=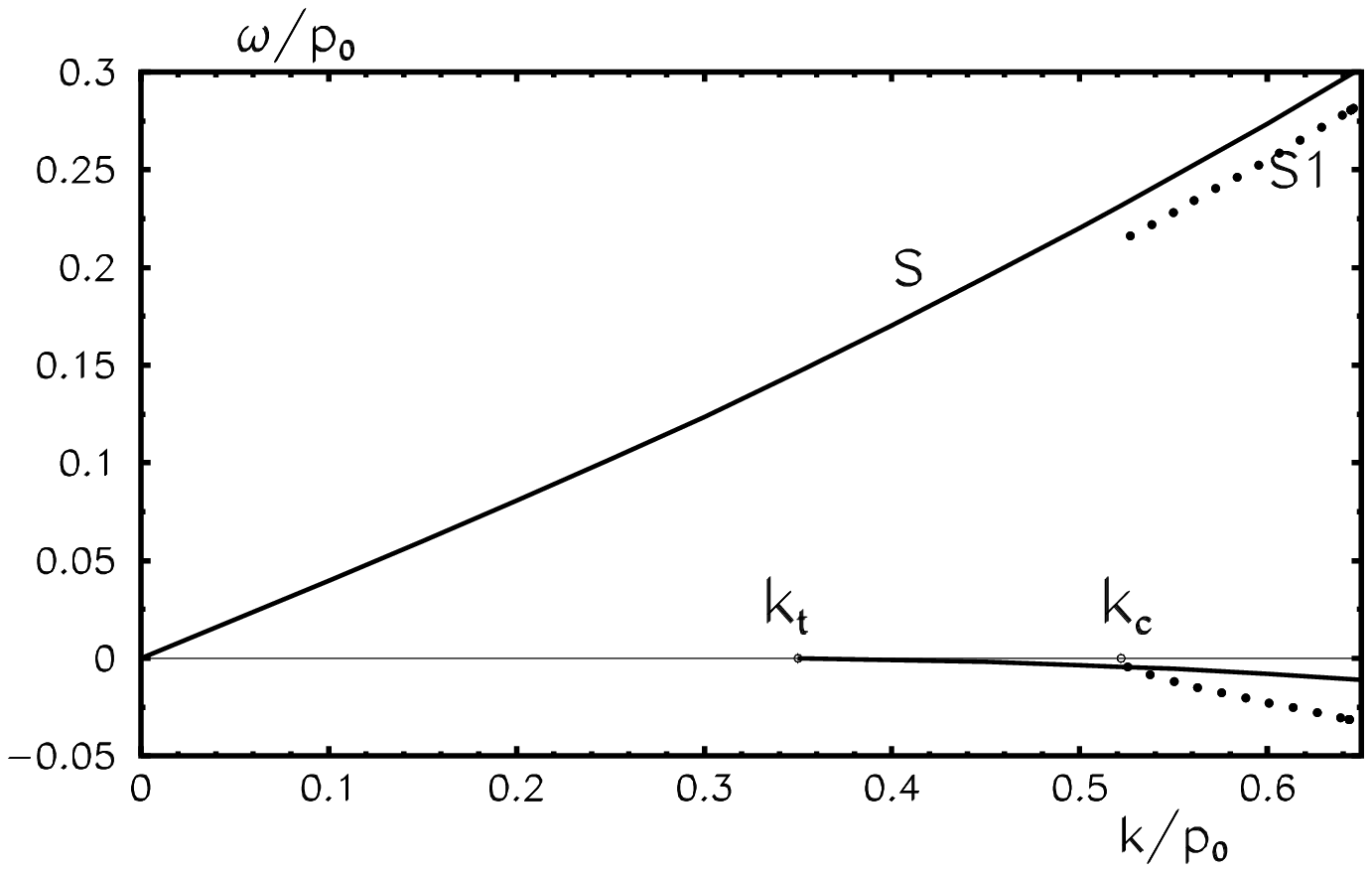,width=7cm} \epsfig{figure=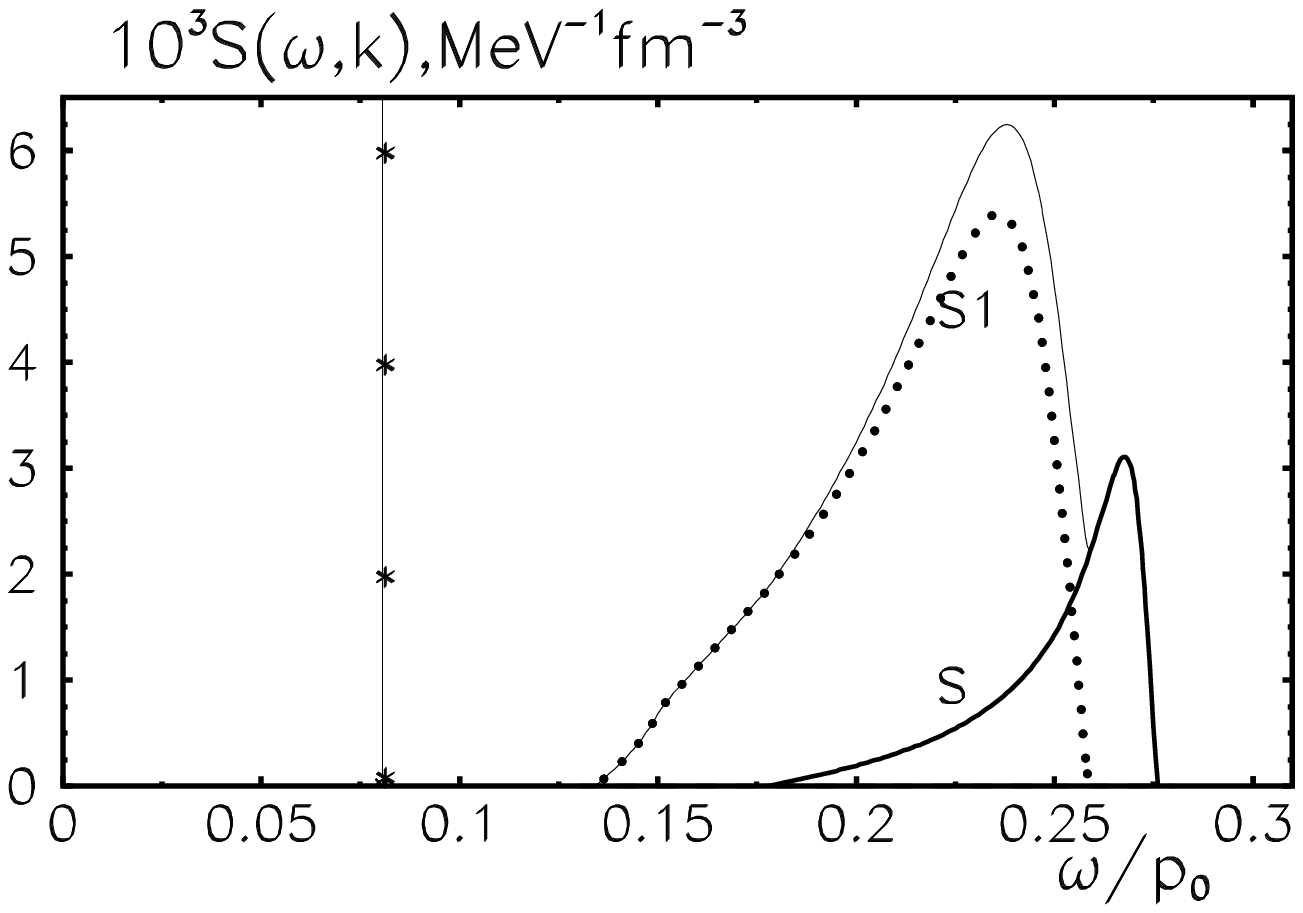,width=7cm}}
\caption{SNM,  $\beta=0.0$. $left$:\, $\omega_{s}(k)$ (solid curve, $S$), $\omega_{s1}(k)$ (dotted curve, $S1$). At $\omega>0$ ($\omega<0$) the real (imaginary) parts of $\omega_{i}(k)$ are demonstrated.
$right$:\, the pole terms of $S_l(\omega,k)$, $l=s,s1$ are shown. The envelope curve $S^e(\omega,k)$ is marked by the thin solid curve for $k/p_0=0.6$ and by the solid with stars for $k/p_0=0.2$.}
\end{figure}

\begin{figure}
\centering{\epsfig{figure=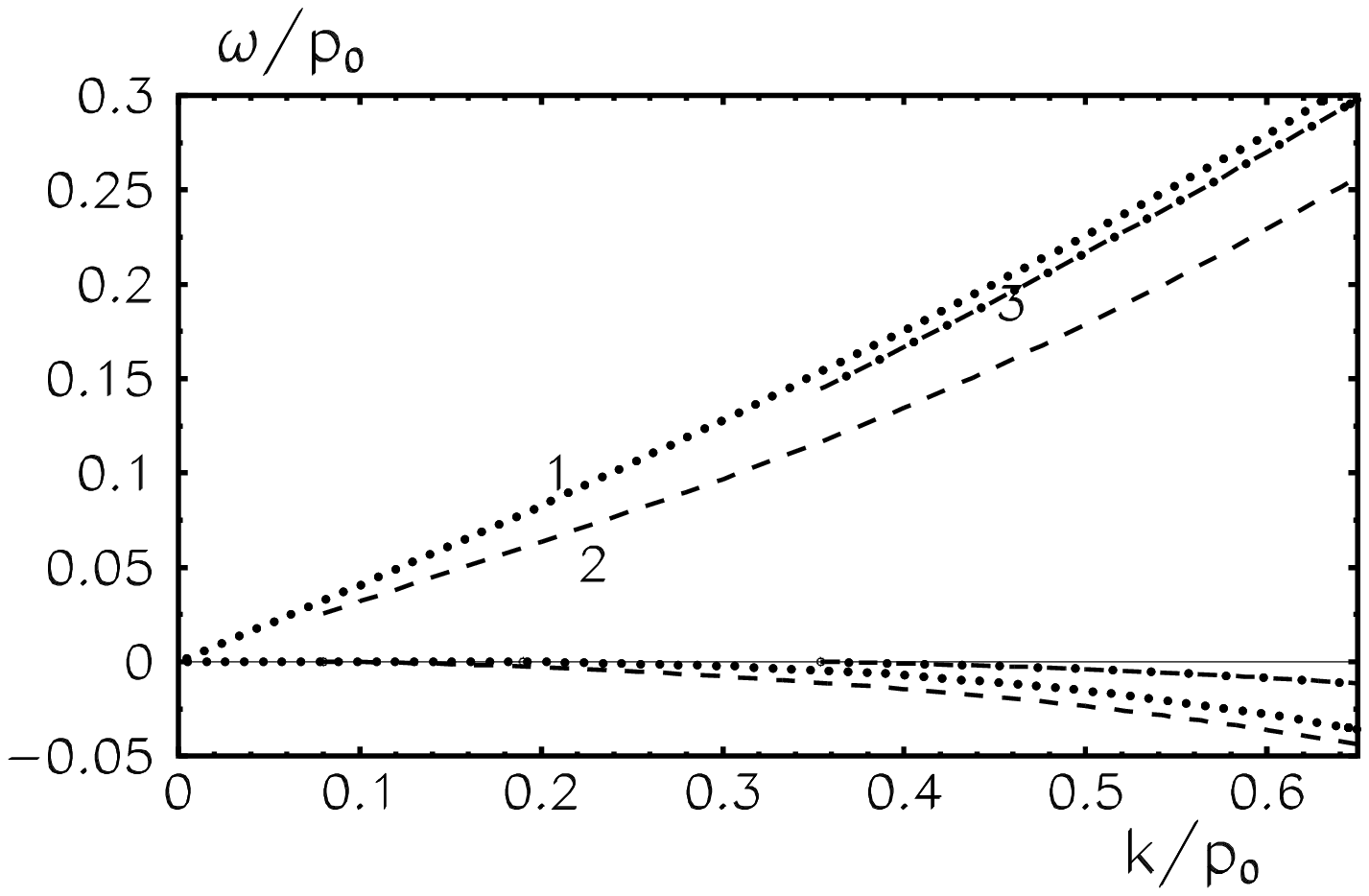,width=7.5cm} \epsfig{figure=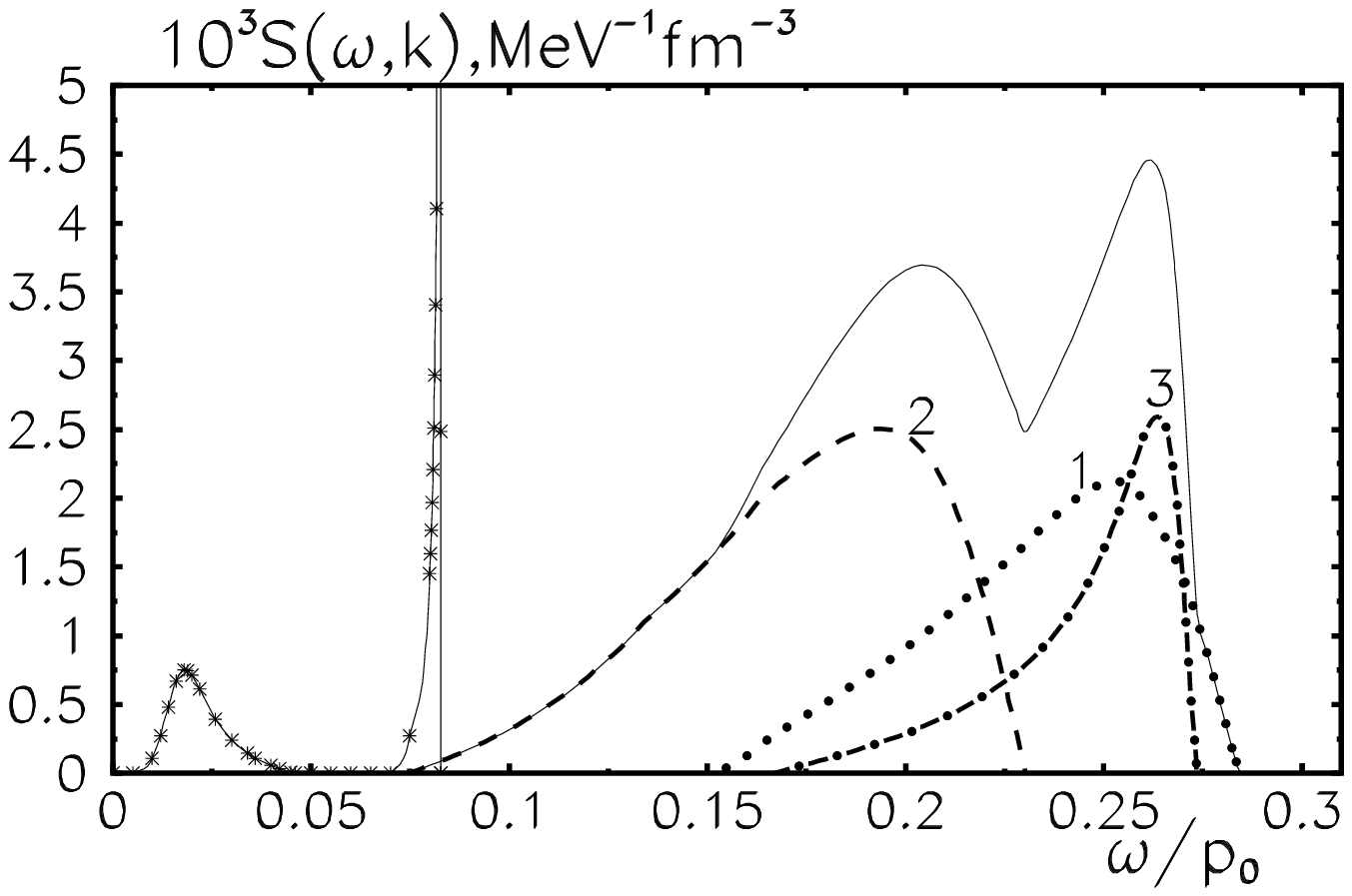,width=7.5cm}}
\caption{ANM, $\beta=0.2$. $left$:\, $\omega_{sn}(k)$ (1, dotted curve), $\omega_{sp}(k)$ (2, dashed curve),   $\omega_{snp}(k)$ (3, dash-dotted). At $\omega>0$ ($\omega<0$) the real (imaginary) parts of $\omega_{si}(k)$ are demonstrated.
$right$:\, the pole terms of $S_l(\omega,k)$, $l=n,p,np$ (numbers $1,2,3$) are shown. The envelope curve $S^e(\omega,k)$ is marked by the thin solid curve for $k/p_0=0.6$ and by the solid with stars for $k/p_0=0.2$.}
\end{figure}

\begin{figure}
\centering{\epsfig{figure=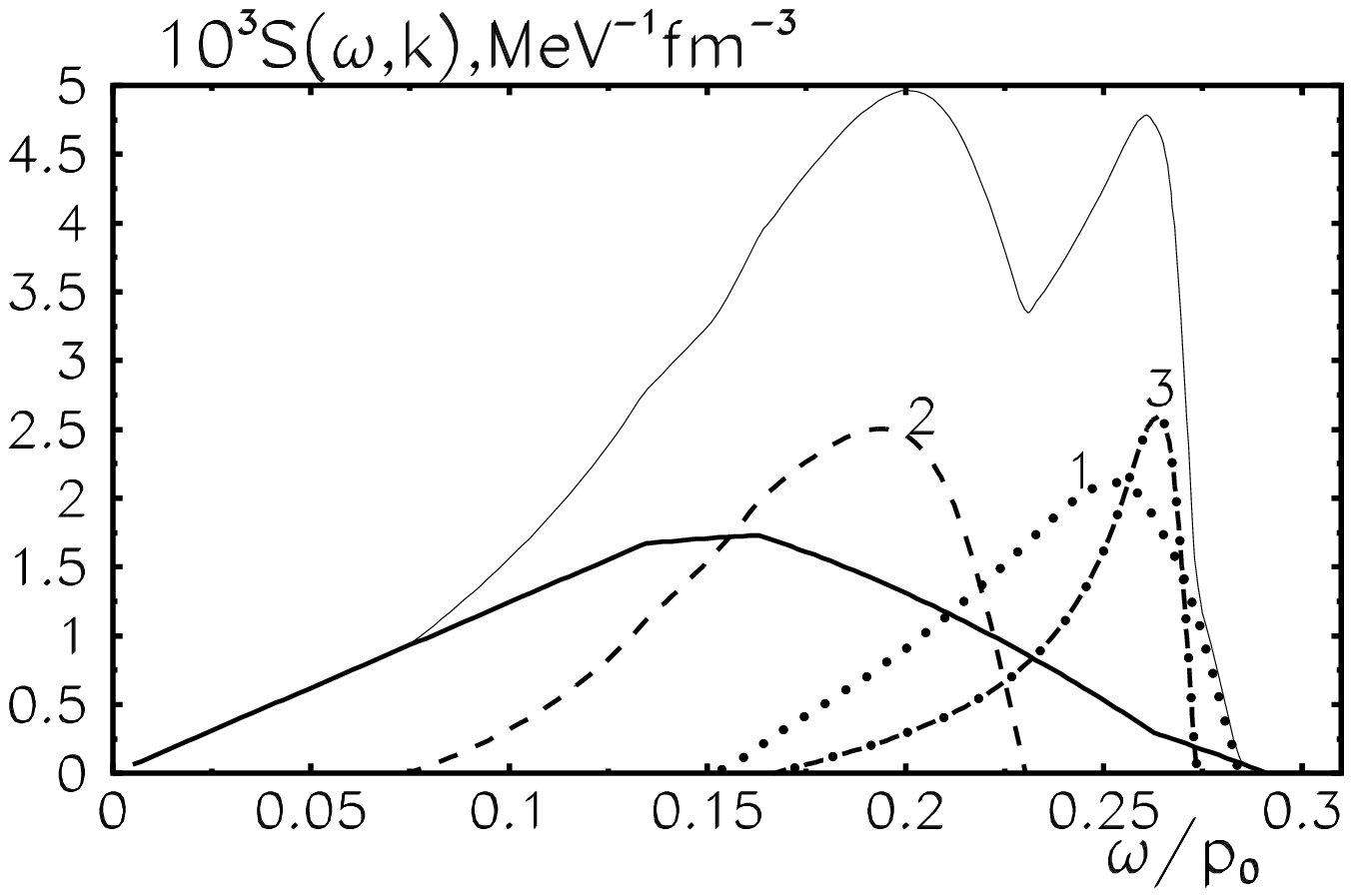,width=7.5cm}\epsfig{figure=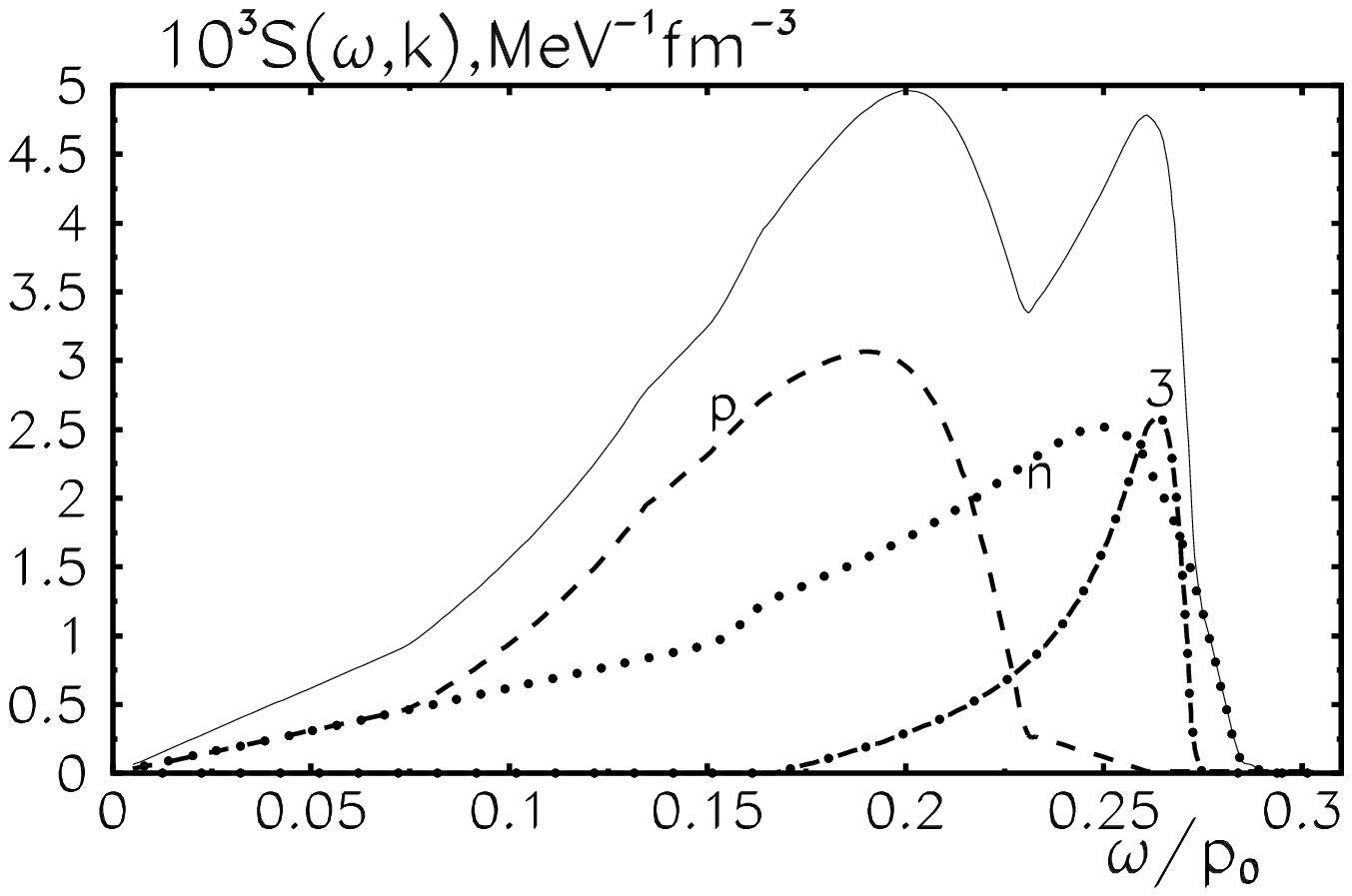,width=7.5cm}}
\caption{ANM, $\beta=0.2$. ($left$): The pole terms, the same as in Fig.2($right$), with additional contribution $S_{fr}(\omega,k)$ (\ref{15S}) due to non-interacting $ph$-pairs (fat solid curve). ($right$): proton and neutron maxima: each of them is presented as the sum of pole and the part of $S_{fr}(\omega,k)$ with corresponding isospin.}
\end{figure}

\end{document}